\documentclass{Interspeech}
\usepackage{multicol}
\usepackage{multirow}

\interspeechcameraready

\title{CHSER: A Dataset and Case Study on Generative Speech Error Correction for Child ASR}

\author[affiliation={1}]{Natarajan}{Balaji Shankar}
\author[affiliation={1}]{Zilai}{Wang}
\author[affiliation={1}]{Kaiyuan}{Zhang}
\author[affiliation={1}]{Mohan}{Shi}
\author[affiliation={1}]{Abeer}{Alwan}

\affiliation{Dept of Electrical and Computer Engineering}{University of California, Los Angeles}{USA}
\email{{balaji1312,zilaiwang2001,kaiyuanzhang,shimohan}@ucla.edu, alwan@ee.ucla.edu}

\usepackage{comment}

\begin{document}

\maketitle

\begin{abstract}
 
 Automatic Speech Recognition (ASR) systems struggle with child speech due to its distinct acoustic and linguistic variability and limited availability of child speech datasets, leading to high transcription error rates. While ASR error correction (AEC) methods have improved adult speech transcription, their effectiveness on child speech remains largely unexplored. To address this, we introduce CHSER, a Generative Speech Error Correction (GenSEC) dataset for child speech, comprising 200K hypothesis-transcription pairs spanning diverse age groups and speaking styles. Results demonstrate that fine-tuning on the CHSER dataset achieves up to a 28.5\% relative WER reduction in a zero-shot setting and a 13.3\% reduction when applied to fine-tuned ASR systems. Additionally, our error analysis reveals that while GenSEC improves substitution and deletion errors, it struggles with insertions and child-specific disfluencies. These findings highlight the potential of GenSEC for improving child ASR.\footnote{This work was supported in part by the NSF.}
\end{abstract}

\section{Introduction}
\label{sec:intro}

Speech Foundation Models (SFMs) have advanced Automatic Speech Recognition (ASR), demonstrating state-of-the-art performance and strong zero-shot generalization \cite{chen2022wavlm, hsu2021hubert, baevski2020wav2vec, Rad23whisper, puvvada2024less}. Their success stems from large-scale pretraining and transformer-based architectures, yet they struggle with child speech, exhibiting significantly high error rates \cite{fan2024benchmarking}. The performance gap between child and adult ASR arises from the distinct characteristics of child speech, including higher pitch, greater intra- and inter-speaker variability, and developmental changes \cite{gerosa2006analyzing, lee1999acoustics, yeung2018difficulties, dutta2022challenges}. Additionally, child speech contains greater disfluencies (e.g., repetitions, restarts), grammatical inconsistencies, and non-standard word usage, which degrade ASR accuracy \cite{zackheim2003childhood}. Variability across speaking styles, such as scripted vs. spontaneous speech, further complicates generalization, necessitating effective post-processing to enhance transcription accuracy.


Traditional ASR error correction (AEC) methods focus on rescoring and hypothesis selection, including n-gram-based rescoring \cite{katz2003estimation}, consensus decoding \cite{mangu2000finding}, and neural rescoring with RNNs and transformers \cite{mikolov2010recurrent, tanaka18_interspeech}. More recent AEC approaches leverage machine translation models for domain adaptation \cite{mani2020}, fine-tuned RoBERTa models for readability improvements \cite{liao2023}, and crossmodal methods incorporating both audio and text representations \cite{lin2023}. Confidence-based correction strategies also integrate utterance rejection for improved robustness \cite{du2022cross}.

Generative speech error correction (GenSEC) has emerged as a promising approach, leveraging large language models (LLMs) to refine ASR outputs by learning error patterns from large-scale paired data. Datasets such as HyPoradise \cite{chen2024hyporadise} and RobustHyPoradise \cite{hu2024large} serve as benchmarks, demonstrating that fine-tuned generative models significantly enhance transcription accuracy. GenSEC exploits full N-best lists for improved hypothesis selection. LoRA-based \cite{hu2021lora} fine-tuning of Llama 2 \cite{touvron2023llama} and Flan T5 \cite{chung2024scaling} models enables efficient adaptation \cite{chen2024hyporadise, la2024flanec} without requiring full retraining. Other approaches include language-space noise embeddings for robustness \cite{hu2024large}, integration of ASR embeddings with LLMs \cite{radhakrishnan2023whispering}, lattice-constrained decoding \cite{ma2024asr}, retrieval-augmented generation for out-of-domain correction \cite{ghosh2024failing}, and multi-expert models for ASR and speech translation \cite{lin2024neko}. However, these approaches are trained on adult speech, limiting their effectiveness for child-specific disfluencies and linguistic patterns.


To address this gap, we introduce CHSER, a GenSEC dataset designed to enhance ASR transcription accuracy for child speech. Our dataset comprises 200k child speech ASR hypothesis pairs across diverse age groups (4–14 years) and speaking styles (scripted, spontaneous), enabling models to learn child-specific error patterns. We evaluate existing GenSEC methods to establish baselines and analyze their limitations in handling child speech variability. Furthermore, we assess the adaptability of child-specific GenSEC models in both zero-shot and fine-tuned settings, demonstrating their impact on improving child ASR performance across different domains.

Our contributions can be summarized as follows:
\begin{itemize}
 \item 
 We introduce the first large-scale dataset for child ASR Generative Speech Error Correction, comprising 200K hypothesis-transcription pairs spanning multiple age groups (4–14 years old) and speaking styles (scripted and spontaneous).
 \item 
Through systematic evaluation, we address four key research questions: \textbf{(RQ1)} The effectiveness of existing GenSEC methods on child speech, \textbf{(RQ2)} The robustness of child-specific error correction models across different domains, \textbf{(RQ3)} The applicability of GenSEC models to fine-tuned ASR systems of differing architectures, and \textbf{(RQ4)} Whether GenSEC alters the nature of ASR errors by analyzing error distribution shifts post correction.\footnote{Our dataset, code, and models are available at \url{https://github.com/balaji1312/CHSER}}
\end{itemize}


\section{CHSER Dataset}

\label{sec:methods}

To construct a benchmark for child speech ASR error correction, we compile a dataset from multiple publicly available child speech corpora, ensuring diversity in age ranges and speaking styles. Our dataset consists of 200K ASR hypothesis-transcription pairs sourced from four major child speech datasets: MyST \cite{pradhan-etal-2024-science}, CMU Kids \cite{eskenazi1997cmu}, CSLU OGI Kids \cite{shobaki2000ogi}, and CHILDES English OCSC Corpus \cite{wagner4846086ohio}. Similar to the procedure used for adult speech in \cite{chen2024hyporadise}, we generated N-best lists of sentence hypotheses using the Whisper-base.en model \cite{Rad23whisper}. To improve hypothesis-transcription pair quality, we apply the following filtering steps: remove repeated utterances, retain only the top 5 hypotheses with the highest probabilities in the N-best list, exclude transcriptions shorter than three words, and discard utterances containing words outside the Whisper tokenizer vocabulary. 
The CHSER dataset in total contains over 200K ASR hypothesis-transcription pairs and is summarized in Table \ref{tab:dataset_statistics}. All dataset divisions into train, dev, and test sets follow the original splits in their respective source datasets. Below, we provide an overview of each source dataset:

The My Science Tutor (MyST) corpus \cite{pradhan-etal-2024-science} is a 400-hour English speech dataset collected from 1,371 students in grades 3–5 (ages 8-11). It features spontaneous, unscripted conversations with a virtual science tutor on topics such as physics, geography, and biology. Of this, we include 64,187 utterances in the train set, 10,458 in the dev set, and 11,591 in the test set.

The CMU Kids Corpus \cite{eskenazi1996kids} consists of read-aloud sentences from 6–11-year-old children. It includes 24 male and 52 female speakers with a total of 5,180 utterances recorded over 9 hours of speech. We incorporate 3,784 utterances in the train set, 642 in the dev set, and 754 in the test set.

The CSLU OGI Kids Corpus \cite{shobaki2000ogi} contains speech from 1,100 children (ages 4–14) across both scripted and spontaneous speech conditions. The read speech section consists of simple words, sentences, and digit strings. The spontaneous speech section includes alphabet recitation and unscripted monologues. We utilize both sections in CHSER, with 18,121 utterances (scripted) and 3,730 utterances (spontaneous) in the train set, 1,971 and 451 in the dev set, and 5,795 and 457 in the test set, respectively.

The OCSC corpus contains 156 hours of speech from 303 primarily monolingual Engish speaking children aged 4–9. The dataset includes a mix of scripted and spontaneous speech, covering structured tasks such as reading passages, and letter and number identification, as well as open-ended tasks like picture descriptions and conversational interactions. As there is no official split, we incorporate 64,581 utterances in the train set, 7,875 in the dev set, and 8,090 in the test set ensuring no speaker overlap between the subsets.

\begin{table}[!htp]
 \centering
 \caption{Statistics of the CHSER dataset split by source, speaking style, and age range. The CHSER dataset consists of hypotheses-transcription pairs generated using Whisper-base.en in a zero-shot beam search setting.}
 \label{tab:dataset_statistics}
 \scriptsize
 \setlength{\tabcolsep}{2pt} 
 \renewcommand{\arraystretch}{1.1}
 \resizebox{\columnwidth}{!}{%
 \begin{tabular}{llc rrr}
 \toprule
 Source & Style & Age Range & Train \# Pairs & Dev \# Pairs & Test \# Pairs \\
 \midrule
 MyST & Spontaneous & 8--11 & 64,187 & 10,458 & 11,591 \\
 CMU Kids& Scripted & 6--11 & 3,784 & 642 & 754 \\
 \multirow{2}{*}{CSLU OGI} & Spontaneous & 4--14 & 3,730 & 451 & 457 \\
 & Scripted & 4--14 & 18,121 & 1,971 & 5,795 \\
 OCSC & Scripted/Spontaneous & 4--9 & 64,581 & 7,875 & 8,090 \\
 \midrule
 \multicolumn{3}{c}{\textbf{All}} & 154,403 & 21,397 & 26,687 \\
 \bottomrule
 \end{tabular}}
\end{table}
\section{Experimental Setup and Results}
\label{sec:results}
\vspace{5pt}
\subsection{RQ1: Do existing GenSEC methods perform well on child speech?}

\begin{table}[!ht]
 \centering
 \caption{WER of Traditional Reranking Models (Trigram LM, Transformer LM) and oracle WER (N-best Oracle ($o_{nb}$), and Compositional Oracle ($o_{cp}$)) on the CHSER dataset. Baseline refers to WER before error correction.}
 \label{tab:classical_models}
 \resizebox{\columnwidth}{!}{%
 \scriptsize
 \begin{tabular}{l rrrrr}
 \toprule
 Dataset & Baseline & Trigram& Transformer & $o_{nb}$ & $o_{cp}$ \\
 \midrule
 All & 30.5 & 29.2 & 29.9 & 17.7 & 8.6 \\
 MyST & 28.1 & 27.0 & 29.9 & 17.2 & 8.1 \\
 CMU Kids & 18.7 & 13.0 & 12.2 & 10.3 & 7.9 \\
 OGI Spontaneous & 40.6 & 36.4 & 39.6 & 25.6 & 10.4 \\
 OGI Scripted & 26.2 & 16.5 & 15.7 & 11.2 & 7.9 \\
 OCSC & 43.1 & 48.9 & 40.2 & 23.3 & 11.2 \\
 \bottomrule
 \end{tabular}}
\end{table}

\begin{table}[!ht]
 \centering
 \caption{WER of LLM-based GenSEC models pre-trained on the HyPoradise adult speech dataset (Llama Hyp, T5 Hyp) and fine-tuned on CHSER (Llama FT, T5 FT) evaluated on the CHSER dataset. Baseline refers to WER before error correction. Bold face numbers indicate best results. Statistically significant results ($p < 0.05$) are indicated with $^{*}$. Baseline definition and notation conventions apply to all subsequent WER tables}
 \label{tab:llm_models}
 \resizebox{\columnwidth}{!}{%
 \scriptsize
 \begin{tabular}{l rrrrr}
 \toprule
 Dataset & Baseline& Llama Hyp & T5 Hyp & Llama FT & T5 FT \\
 \midrule
 All & 30.5 & 31.0 &27.3& 24.8& \textbf{21.8}$^{*}$ \\
 MyST & 28.1 & 27.2 & 26.4& 22.5& \textbf{20.7}$^{*}$ \\
 CMU Kids& 18.7 & 16.6 & 15.5& 13.6& \textbf{12.9}$^{*}$ \\
 OGI Spontaneous & 40.6 & 37.6 & 36.3& 32.8& \textbf{29.7}$^{*}$ \\
 OGI Scripted & 26.2 & 30.4 & 26.1& 16.2& \textbf{13.3}$^{*}$ \\
 OCSC & 43.1 & 47.9 & 40.7& 40.6& \textbf{32.2}$^{*}$ \\
 \bottomrule
 \end{tabular}}
\end{table}

\begin{table}[!ht]
 \centering
 \caption{WER of In-Context Learning with GPT-4o mini with different prompting examples (n=0, n=1, n=5) on the CHSER dataset.}
 \label{tab:gpt4o_prompting}
 \scriptsize
 \begin{tabular}{l rrrrr}
 \toprule
 Dataset & Baseline & n=0& n=1 & n=5 \\
 \midrule
 All & 30.5 & 42.1 & \textbf{27.3} & 27.4 \\
 MyST & 28.1 & 34.1 & 25.5 & \textbf{25.0}\\
 CMU Kids& 18.7 & 28.9 & \textbf{13.8}$^{*}$ & 16.4 \\
 OGI Spontaneous & 40.6 & 55.5 & 38.3 & \textbf{35.7}$^{*}$ \\
 OGI Scripted & 26.2 & 40.0 & \textbf{19.6}$^{*}$ & 23.0 \\
 OCSC & 43.1 & 77.0 & \textbf{39.6}$^{*}$ & 40.2 \\
 \bottomrule
 \end{tabular}
\end{table}
To assess the effectiveness of Generative Speech Error Correction (GenSEC) methods on child speech, we evaluate commonly used approaches including Llama 2 \cite{touvron2023llama}, Flan T5 \cite{chung2024scaling}, and In-Context Learning (ICL) using baselines established in \cite{yang2023generative, chen2024hyporadise}. We calculate WER and assess statistical significance using the sctk package from the NIST Scoring Toolkit \cite{NIST-SCTK} to ensure a fair comparison across methods.

\begin{itemize}
 \item 
 \textbf{Traditional Reranking-Based Models:}
 
 We include two language model baselines that perform log-probability score-based error correction:
 (i) Trigram LM: A trigram language model trained on in-domain textual data from the CHSER train set, serving as a strong traditional baseline. 
 (ii) Transformer LM: A transformer-based language model pre-trained on the LibriSpeech LM corpus \cite{PanayotovCPK15}, then fine-tuned on in-domain text from the CHSER train set. 
 The effectiveness of these models in correcting errors is reported in Table \ref{tab:classical_models}. Additionally, we report two oracle WER values as theoretical performance upper bounds: 
(i) Oracle-NB ($o_{nb}$): The WER of the best candidate in the N-best hypothesis list.
(ii) Oracle-CP ($o_{cp}$): The lowest possible WER achievable based on an optimal composition of tokens across multiple hypotheses. 
 \item 
 \textbf{Fine-Tuned GenSEC Models:}
 
 To evaluate the generalization of GenSEC models to child speech, we assess Flan T5 large \cite{chung2024scaling} and Llama 2 7B \cite{touvron2023llama} models pre-trained on HyPoradise \cite{chen2024hyporadise}, a dataset designed for adult speech correction. Additionally, we instruction fine-tune Flan T5 large and Llama 2 7B models on CHSER using 8-bit quantization \cite{jacob2018quantization} with LoRA adapters \cite{hu2021lora} for 5 epochs. The results for these models are presented in Table \ref{tab:llm_models}.
 \item 
 \textbf{In-Context Learning (ICL) with GPT-4o mini:}
 
 We investigate In-Context Learning (ICL) \cite{xie2022an} as an alternative to fine-tuning, evaluating GPT-4o mini \cite{hurst2024gpt} for error correction. Three prompting strategies are tested:
 (i) n=0 (Zero-shot prompting), where the model is provided only the ASR output and no examples; 
 (ii) n=1 (One-shot prompting), where one correction example is provided; and 
 (iii) n=5 (Few-shot prompting), where five examples are provided before generating corrections. 
 The performance of GPT-4o mini across these strategies is shown in Table \ref{tab:gpt4o_prompting}.
\end{itemize}

Our findings, summarized in Tables \ref{tab:classical_models}, \ref{tab:llm_models}, and \ref{tab:gpt4o_prompting}, demonstrate key trends in child speech error correction, highlighting the strengths and limitations of different approaches. 

As shown in Table \ref{tab:classical_models}, traditional reranking-based models (Trigram LM and Transformer LM) provide only minor WER improvements over the baseline ASR model. The Trigram LM performs slightly better than the Transformer LM across most datasets. However, these methods fail to generalize effectively, with limited impact on spontaneous speech datasets such as OGI Spontaneous and OCSC, where the WER remains significantly high. Oracle analyses suggest that optimal hypothesis selection and compositional error correction could yield substantial additional improvements, though no current method reaches these theoretical bounds. 

Table \ref{tab:llm_models} demonstrates that fine-tuned LLM-based GenSEC models (T5 FT, Llama FT) significantly outperform both methods in table \ref{tab:classical_models}, and pre-trained models. The Flan T5 model finetuned on the CHSER dataset (T5 FT) achieves the lowest WER across all datasets, with a relative WER reduction of 28.5\% over the baseline ASR model. Notably, T5 FT reduces absolute WER by 12.9\% on OGI Scripted, one of the largest improvements observed, demonstrating that structured read speech benefits the most from fine-tuning.
Performance remains weaker on spontaneous datasets, such as OCSC, where WER remains at 32.2\%, indicating that modeling child speech variability remains a challenge.

Table \ref{tab:gpt4o_prompting} presents the WER results of ICL-based correction using GPT-4o mini, tested with different prompting strategies (zero-shot, one-shot, few-shot). Zero-shot prompting (n=0) performs significantly worse than all other methods, often increasing WER rather than reducing it. One-shot and Few-shot prompting improves on the baseline, but are still weaker than fine-tuned models, demonstrating that ICL, while useful, is not a substitute for fine-tuning when domain adaptation is necessary.

\subsection{RQ2: Are child error correction models robust across different child specific domains?}
 
\begin{table}[!ht]
 \centering
 \caption{WER across T5 Fine-Tuned on CHSER (T5 FT CHSER), and T5 Fine-Tuned on MyST subset of CHSER (T5 FT MyST), on the CHSER dataset.}
 \label{tab:wer_comparison}
 \scriptsize
 \begin{tabular}{l rrr}
 \toprule
 Dataset & Baseline & T5 FT CHSER & T5 FT MyST \\
 \midrule
 All & 30.5 & \textbf{21.8}$^{*}$ & 22.4 \\
 MyST & 28.1 & 20.7 & \textbf{19.9}$^{*}$ \\
 CMU Kids& 18.7 & \textbf{12.9}$^{*}$ & 16.2 \\
 OGI Spontaneous & 40.6 & 29.7 & \textbf{29.2} \\
 OGI Scripted & 26.2 & \textbf{13.3}$^{*}$ & 17.7 \\
 OCSC & 43.1 & \textbf{32.2}$^{*}$ & 35.7 \\
 \bottomrule
 \end{tabular}
\end{table}
To assess the effectiveness of GenSEC models on child speech, we examine whether a model trained on one subset of the CHSER dataset can generalize to other domains. Specifically, we compare the WER of three systems in Table \ref{tab:wer_comparison}:
A baseline ASR model, which provides the uncorrected transcriptions (Baseline),
A T5 model fine-tuned on the entire CHSER dataset (T5 FT CHSER),
A T5 model fine-tuned only on the MyST subset of CHSER (T5 FT MyST). All T5 models are instruction fine-tuned in settings similar to those in Section 3.1.

Our results indicate that T5 FT MyST performs well on its in-domain data (MyST) but struggles to generalize to other scripted datasets, such as CMU Kids and OGI Scripted, where its WER increases significantly. This suggests that models fine-tuned on a single child speech domain capture domain-specific characteristics effectively but lack the adaptability needed for diverse datasets.

In contrast, T5 FT CHSER consistently achieves lower WER across most domains, demonstrating that training on a diverse dataset improves generalization. These results highlight the importance of multi-domain training for GenSEC models, as fine-tuning on a single dataset may lead to overfitting and reduced effectiveness in unseen child speech domains.

\subsection{RQ3: Do error correction models offer robustness for fine-tuned ASR models of different architectures?}
\begin{table}[!ht]
 \centering
 \caption{WER reduction with a T5 model fine-tuned on the MyST subset of CHSER, evaluated across different ASR models on the MyST test set. The ASR hypotheses are generated using versions of Whisper-tiny.en (Whisper Tiny), Whisper-small.en (Whisper Small), and WavLM-large (WavLM) fine-tuned on the MyST train set.}
 \label{tab:t5_correction}
 \scriptsize
 \begin{tabular}{l rr}
 \toprule
 ASR Model & Baseline & T5 Corrected\\
 \midrule
 WavLM & 10.1 & \textbf{9.2}$^{*}$ \\
 Whisper Small & 11.4 & \textbf{10.3}$^{*}$ \\
 Whisper Tiny & 12.8 & \textbf{11.1}$^{*}$ \\
 \bottomrule
 \end{tabular}
\end{table}

Since the CHSER dataset is generated from beam search hypotheses of a zero-shot Whisper-base.en model, an important question is whether WER improvements from error correction generalize to different ASR models, including those fine-tuned on child speech.

To evaluate this, as an example we test a Flan T5 model fine-tuned on the MyST subset of CHSER (instruction fine-tuned in settings similar to those in Section 3.1) against transcriptions from three ASR models, as shown in Table \ref{tab:t5_correction}: Whisper-tiny.en (attention encoder decoder (AED), 39M parameters, pretrained on 680k hours), Whisper-small.en (AED, 244M parameters, pretrained on 680k hours), and WavLM-large (CTC encoder-only model, 316M parameters, pretrained on 94k hours), all fine-tuned on the MyST dataset. We note that our reported WER for the beam search decoding of Whisper-based models is higher than the WER obtained via greedy decoding as reported in the literature \cite{fan2024benchmarking, attia2024kid}, likely due to the additional temperature setting applied during beam search to encourage more diverse output generation.

The results, presented in Table \ref{tab:t5_correction}, demonstrate that the Flan T5 model fine-tuned on the MyST section of the CHSER dataset provides consistent WER (9-13.3\%) reductions across all fine-tuned ASR models, demonstrating its robustness in improving transcription accuracy irrespective of the underlying ASR model.

\subsection{RQ4: Are the errors present after error correction different from errors before correction?}

While GenSEC improves transcription accuracy, it is equally important to examine whether the nature of transcription errors shifts post-correction. We categorize errors in the original CHSER dataset and those remaining after correction using the best performing model from Section 3.1 (T5 FT), and conduct the following analysis:
\begin{itemize}
\item
\textbf{Error Type Distribution:}

We analyze errors using the insertion, substitution, and deletion (I/S/D) framework to determine whether the fine-tuned T5 model introduce biases toward specific types of corrections. Figure \ref{fig:error_type_analysis_combined} shows that substitutions are the most frequent error type both before and after correction. While GenSEC significantly reduces substitutions and deletions, it struggles to correct insertion-based errors, suggesting that erroneous insertions in ASR outputs are more challenging to remove than missing or misrecognized words.
\begin{figure}[!ht]
 \centering
 \includegraphics[width=0.63\columnwidth]{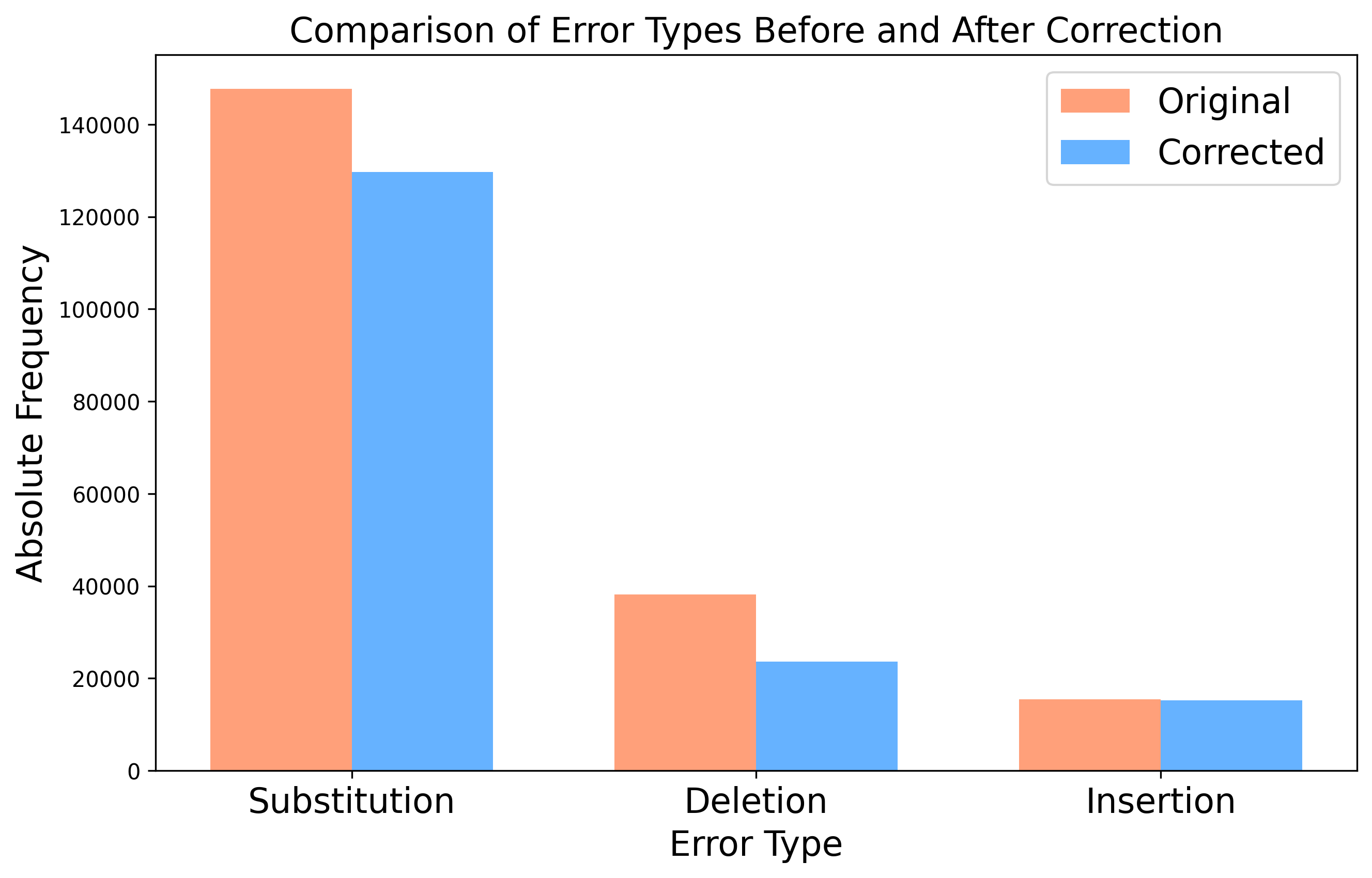}
 \caption{Comparison of error types (substitutions, deletions, insertions) before and after error correction. Errors are categorized by comparing ASR hypotheses to ground truth transcriptions.}
 \label{fig:error_type_analysis_combined}
\end{figure}
\begin{figure}[!ht]
 \centering
 \includegraphics[width=0.63\columnwidth]{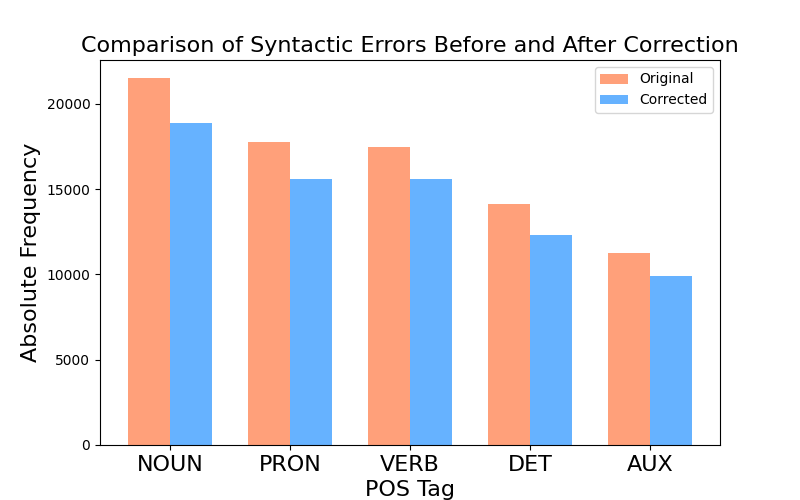}
 \caption{Syntactic errors categorized by part-of-speech (POS) tags (NOUN, PRON, VERB, DET, AUX) before and after error correction. Errors are identified by comparing ASR hypotheses to ground truth transcriptions.}
 \label{fig:syntactic_errors_comparison}
\end{figure}
\begin{figure}[!ht]
 \centering
 \includegraphics[width=0.63\columnwidth]{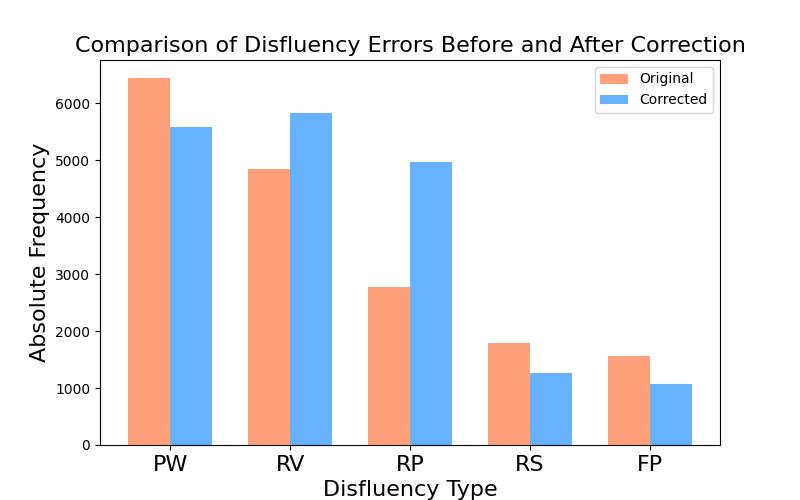}
 \caption{Disfluency errors - filled pauses (FP), partial words (PW), repetitions (RP), revisions (RV), and restarts (RS), before and after error correction. Disfluencies are identified in the ground truth transcriptions, ASR hypotheses, and corrected outputs to track correction rates.}
 \label{fig:disfluency_errors_comparison}
\end{figure}

\item
\textbf{Syntactic Errors:}

To evaluate whether the fine-tuned T5 model introduces syntactic shifts, we categorize substitution-based errors by part-of-speech (POS) tags using spaCy \cite{spacy2020}. This analysis is conducted by aligning the ground truth transcription to the hypothesis pre- and post-correction, and focuses specifically on words that were substituted. The results, shown in Figure \ref{fig:syntactic_errors_comparison}, reveal a consistent reduction in syntactic errors across the top five most frequent POS categories.

\item
\textbf{Disfluency Errors:}

Disfluencies are a common feature of child speech, and we analyze their detection and correction by first identifying the number of disfluencies present in the ground truth transcripts. We then compare this with the number detected in the ASR hypotheses, tracking the detection error rate for each category: filled pauses (FP), partial words (PW), repetitions (RP), revisions (RV), and restarts (RS) using the procedure detailed in \cite{romana2024automatic}. Finally, we evaluate how these errors change post-correction to determine whether GenSEC models mitigate or amplify disfluency related errors. Figure \ref{fig:disfluency_errors_comparison} shows that while the fine-tuned T5 model improves the detection rate for partial words, filled pauses and restarts, it struggles with repetitions and revisions, leading to an increase in their error rates after correction. This suggests that GenSEC models primarily focus on syntactic and lexical refinements but remain challenged by disfluency recognition, potentially misclassifying them or introducing new errors.

\end{itemize}
\section{Conclusion}
\label{sec:conclusion}
This study presents CHSER, a Generative Speech Error Correction (GenSEC) dataset comprising 200K hypothesis-transcription pairs spanning diverse age groups and speaking styles specifically designed to address the challenges of ASR transcription for child speech. Through a comprehensive evaluation, we systematically investigated four key research questions:
\begin{itemize}
 \item
(RQ1) Effectiveness of existing GenSEC methods: We found that pre-trained GenSEC models (LLaMA 2, Flan T5), originally developed for adult speech, struggle to generalize to child speech, highlighting the need for child-specific adaptation. Fine-tuning on CHSER significantly improved WER across multiple child speech datasets, with a fine-tuned Flan T5 model achieving a 28.5\% relative WER reduction in a zero-shot setting, demonstrating better generalization across diverse domains.
\item 
(RQ2) Robustness of child-specific correction models across domains: Models fine-tuned on a single dataset perform well in-domain but struggle to generalize, especially to datasets with different speaking styles.
\item 
(RQ3) Applicability to fine-tuned ASR models: When applied to fine-tuned ASR systems with differing architectures, GenSEC remained effective, achieving a 13.3\% relative WER reduction.
\item 
(RQ4) Shifts in error distribution post-correction: Our error analysis revealed that while GenSEC significantly reduces substitutions and deletions, it struggles with insertion errors and child-specific disfluencies such as repetitions and revisions, underscoring the linguistic challenges unique to child ASR.
\end{itemize}
These findings underscore the potential of GenSEC to improve the performance of child ASR. Future directions include exploring hybrid approaches that integrate language modeling with prosodic and phonetic cues to improve GenSEC’s handling of disfluencies present in child speech.

\bibliographystyle{IEEEtran}
\bibliography{mybib}

\end{document}